\documentclass[aps,groupedaddress,preprint]{revtex4}
\usepackage{color}
\usepackage{epsf}
\usepackage{epsfig}
\usepackage{epstopdf}
\usepackage{graphicx}
\usepackage{latexsym}
\usepackage{amssymb}
\usepackage{url}

\begin{document}
%\draft

%
\begin{flushleft}
{\em Phys. Rev. A 89, 032702 (2014)}
\newline
\end{flushleft}

\title {Search for isotope effects in projectile and target ionization in swift He$^+$ on H$_2$ / D$_2$ collisions}

\author{F.~Trinter$^1$}%
\email{trinter@atom.uni-frankfurt.de}
\author{M.~Waitz$^1$}%
\author{M.~S.~Sch\"offler$^1$}%
\author{H.-K.~Kim$^1$}%
\author{J.~Titze$^1$}%
\author{O.~Jagutzki$^1$}%
\author{A.~Czasch$^1$}%
\author{L.~Ph.~H.~Schmidt$^1$}%
\author{H.~Schmidt-B\"ocking$^1$}%
\author{R.~D\"orner$^1$}%

\affiliation{$^1$ Institut f\" ur Kernphysik, Goethe-Universit\" at
Frankfurt am Main, 60438 Frankfurt, Germany}%

\date{\today}

\begin{abstract}

Using the COLTRIMS technique we have measured the simultaneous projectile and target ionization in collisions of He$^+$ projectiles with a mixture of gaseous H$_2$ and D$_2$ for an incident projectile energy of 650 keV. Motivated by Cooper et al. [Phys. Rev. Lett. \textbf{100}, 043204 (2008)] we look for differences in the ionization cross section of the two isotopes with highest resolution and statistical significance. Contributions of the electron-electron and the electron-nucleus interactions have been clearly separated kinematically by measuring the longitudinal and transverse momentum of the recoiling ion. We find no significant isotope effect in any of our momentum distributions.

\end{abstract}


\maketitle

{\bf I. Introduction} \vskip 3mm

On the atomic scale, interaction such as
photo ionization and charged particle impact is
completely dominated by the electromagnetic
force. If one neglects the tiny hyperfine
structure splitting, the nucleus just provides
the Coulomb potential in which the electrons
move. The ionization of single atoms by charged
particles is therefore not expected to depend on
the isotope. Because the chemical behavior of an
atom is largely determined by its electronic
structure, different isotopes also exhibit very
similar chemical behavior. The main exception to
this is the kinetic isotope effect: due to their
larger masses, heavier isotopes tend to react
slightly more slowly than lighter istotopes of
the same element. This effect is most pronounced
for H and D, because deuterium has twice the mass
of the proton. The mass effect between H and D
also affects the behavior of their respective
chemical bonds by changing the center of gravity
(reduced mass) of the atomic system \cite{Muc71,
Ber73}. For heavier elements the mass-difference
effects on chemistry are usually negligible. In
general, absolute and differential cross sections
for ionization by fast charged particles are
believed to be isotope independent.

In surprising contrast, Cooper and co-workers
observed differences in quasi-elastic (energy
transfer is small compared to the incident energy
of the scattered particles) electron-scattering
cross sections from gaseous H$_2$, D$_2$ and HD
molecules \cite{Coo08}. Electron scattering with
a relatively high momentum transfer was measured
and experimentally determined cross sections were
compared with calculated ones using Rutherford
scattering theory, where the scattering cross
sections depend only on the charge and not on the
mass of the target. Cooper et al. found however,
that the ratio I(H)/I(D) of the scattering
intensities showed a much smaller value than
expected from this conventional theory. For a
50:50 H$_2$-D$_2$ gas mixture they found a ratio
of the respective scattering intensities of
$\approx$ 0.7 (a shortfall of $\approx$ 30 $\%$),
while for the HD gas $\approx$ 1 has been seen as
expected. Following these results, the absolute elastic scattering
probability for H$_2$ and the appropriate
geometrical cross section seems to be smaller
than expected. On the other hand, the ratio of
the electron scattering intensities of H and D in the case
of the HD gas agrees well with the predictions of
Rutherford scattering.

Cooper et al. \cite{Coo08} speculated that
entanglement between identical nuclei (in
H$_2$ and D$_2$) could play a crucial role. They
argued that possibly short-time entanglement of
the protons with their adjacent electrons and
nuclei might reduce the cross section. With this
speculation they challenged the common independent
scatterer model. They assumed that for fast,
large momentum transfer electron scattering on a
very short time scale the H$_2$ wave function can
not be separated to a product of single particle
wave function for electrons and nuclei and that
the scattering is not effectively the sum of the
scattering at any of these isolated particles.
The entangled scatterer could lead to a smaller
cross section in contrast to conventional theory.
Then the lifetime of these quantum entanglements
could be measured indirectly. They estimate that
their scattering processes correspond to typical
scattering times of about 500 attoseconds. If the
entanglement between neighboring particles is
responsible for the anomaly in cross sections,
the lifetime of the entanglement must be much
longer than the scattering time. In this case,
the lifetime of the entanglement could be probed
by varying the momentum transfer and therefore
the scattering time.

Related anomalies in the cross sections of
hydrogen have been observed for different
scattering systems: neutron scattering from
H$_2$O and D$_2$O molecules \cite{Cha97},
electron and neutron scattering from solid
polymers \cite{Cha03}, neutron scattering from
molecular hydrogen \cite{Cha05}. All these
results represent a challenge for conventional
scattering theory as well as for molecular
spectroscopy. No quantitative theory so far could
explain all these anomalies, but all suggested
approaches today contain quantum entanglement \cite{Cha05, Gid05, Rei05, Kar00, Kar02, Cha052}. There is also criticism of the measurements by Cooper et al., see \cite{Moreh} and references therein. Moreh suspects that the origin of the above deviations is instrumental and not due to any real deviation from the Rutherford formula. He argues that the heavier gas component in any binary gas mixture from an inlet spends on average more time in the interaction region which would cause the observed anomaly.

\vskip 3mm

To shine more light on these unexpected
experimental findings in electron and neutron
scattering, we here use a different approach. We
investigate the following reaction:

\begin{eqnarray}
		He^+ + H_2 / D_2 \rightarrow He^{2+} + {H_2}^+ / {D_2}^+ + 2e^-
\end{eqnarray}

In this process simultaneously one electron is
ejected from the target (H$_2$ / D$_2$) and one from the
projectile (He$^+$). There are two mechanisms,
termed $ee$ and $ne$ which contribute to this
reaction. In the $ee$ process the projectile
electron is knocked out by an interaction with
the target electron. This violent collision
between two bound electrons leads to ejection of
both of them from their respective binding. In
the second process, termed $ne$, the projectile
electron is ejected by an interaction with the
target nucleus and a second, independent
interaction between the projectile nucleus and
one of the target electrons facilitates the
ejection of the target electron. Depending on the
impact energy one of these two processes
dominates \cite{Doe94}. The $ee$ process occurs
at rather large impact parameters. The electrons
knock  each other out, while the nuclei are
spectators in the reaction (see Fig. 1a).
Therefore the overall momentum transfer to the
target nucleus is small. A projectile velocity
greater than 2 a.u. is needed to initiate the
process. This threshold is due to the fact that
the target electron, as seen from the projectile
frame, must have sufficient kinetic energy that
it can ionize the projectile and simultaneously
escape from the target. For He a projectile energy of 0.4 MeV is equivalent to an electron energy of 54 eV, which is the projectile binding energy \cite{Doe94, Mon921}. For the $ne$ process on the
contrary the impact parameters are much smaller
leading to a larger transverse momentum transfer (perpendicular to the projectile beam direction)
and in addition the longitudinal momentum (parallel to the projectile beam direction) is also
compensated by the target nucleus \cite{Mon922}. Both electron loss processes are therefore
separated in momentum space of the recoiling
molecular hydrogen ion. Fig. 1c shows the recoil
ion momentum distribution for a collision of
$He^+ + He \rightarrow He^{2+} + He^+ + 2e^-$ at
1 MeV. The most detailed investigations also measured the momenta
of both electrons involved in the loss processes \cite{Kol02, Ferger05, Wang11}.

The key idea of our experiment is, that the $ne$
mechanism involves a violent scattering of the
electron bound to the He$^+$ at the nucleus of
the H$_2$ or D$_2$, while the $ee$ mechanism involves only the electrons. The 30 \% isotope differences
in the scattering of free electrons at the nuclei reported in
\cite{Coo08} might thus also affect the
scattering of bound electrons in our collision
system. The strength of our experiment is, that
the $ee$ mechanism allows for robust in situ
normalization of the data. While in any electron
or neutron scattering experiment the comparison
of cross sections for the different isotopes
relies on the knowledge of the isotope ratio in
the target gas mixture, this possible source
of systematical error is not present in our
experiment.
The $ee$ mechanism in our case is a
scattering between the electrons and the nuclei
are not active participants. Hence the previously
reported isotope effects in electron and neutron
scattering would not influence the $ee$
contribution to the electron loss channel but
only the $ne$ contribution. We therefore search for
differences in the $ne$ versus $ee$ contributions
between the isotopes. While no standard
scattering theory would predict such an isotope
effect the recent results by Cooper et al. on free
electron scattering would suggest such a
difference of up to 30 \%.

\begin {figure}[htbp]
  \begin{center}
    \includegraphics[width=1.0\linewidth]{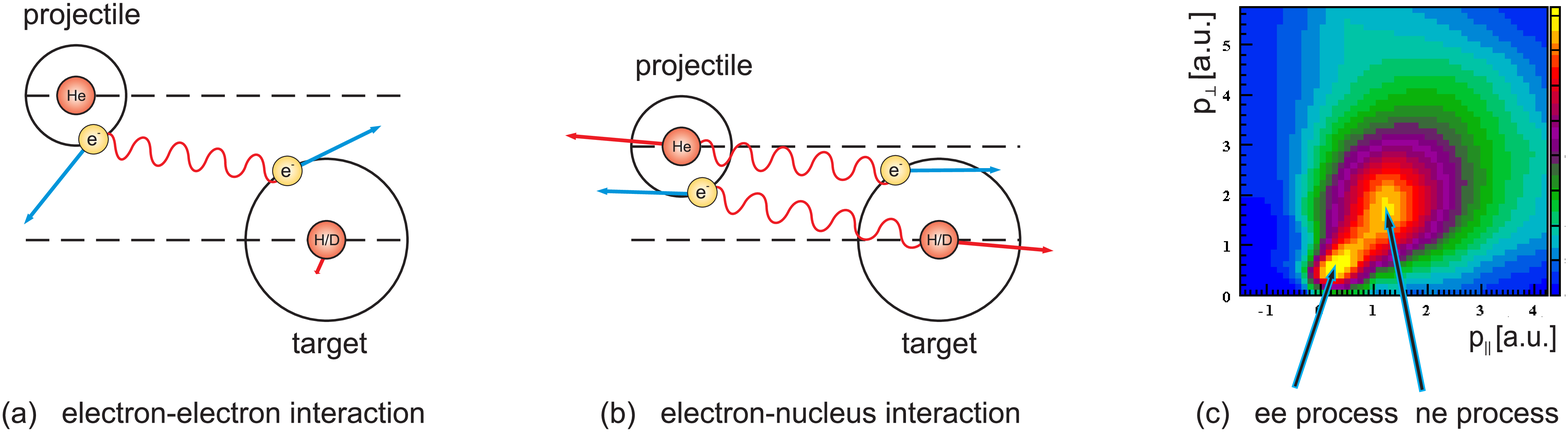}
    \caption{(Color online) COLTRIMS allows to separate experimentally the contribution of $ee$ interaction from the $ne$ interaction to the ionization of the projectile as shown in the doubly differential cross section $d^2\sigma/(dp_{\parallel}dp_{\perp})$. Fig. 1c for illustration shows the recoil ion momentum distribution for a collision of $He^+ + He \rightarrow He^{2+} + He^+ + 2e^-$ at 1 MeV, the horizontal axis shows the longitudinal recoil ion momentum (parallel to the projectile beam direction) and the vertical axis the transverse recoil ion momentum (perpendicular to the projectile beam direction).}
  \label{fig1}
  \end{center}
\end {figure}

\vskip 10mm
{\bf II. Experiment}
\vskip 3mm

In our experiment we use the COLTRIMS (Cold Target Recoil Ion Momentum Spectroscopy) technique \cite{Doe00, Ull03, Jah04} to examine the electron loss process
\cite{Doe951, Doe952, Wu94, Wu97} in collisions
of He$^+$ projectiles with a gas mixture of 50
$\%$ H$_2$ and 50 $\%$ D$_2$. The He$^+$ ion beam
is provided by the 2.5 MV Van-de-Graaff
accelerator at the Institut f\"ur Kernphysik of
the Goethe-Universit\"at in Frankfurt. The beam
is collimated using three sets of adjustable
slits. In front of the reaction zone an
electrostatic deflector (beam cleaner) is used to
separate the He$^+$ beam from charge state
impurities (He$^0$ or He$^{2+}$). The reaction
takes place in the overlap region of the He$^+$
beam with a supersonic gas jet providing the
H$_2$ and D$_2$. An electric field projects the
produced recoil ions (H${_2}^+$ / D${_2}^+$)
towards a time- and position-sensitive detector
\cite{Roe13, Jagutzki} yielding an acceptance angle of
4$\pi$ for ions up to 5 a.u. momentum. Measuring
the impact position on the detector and its
time-of-flight, the particle trajectories and
thus the particle momenta can be determined. To
optimize the ion momentum resolution, an
electrostatic lens is incorporated into the
spectrometer system. Trajectory calculations
including such lenses can be found in
\cite{Sch11}. The diminishing influence of the
extended interaction region on the momentum
resolution can be strongly reduced by this
focusing geometry. A drift tube following the
acceleration part of the spectrometer yields
focusing of the times-of-flight also in the
third direction. The spectrometer has an
electrical field of 6.5 V/cm and a length of 28
cm plus 150 cm recoil drift. Downstream of the
spectrometer another electrostatic deflector is
used to charge state analyze the projectile beam
after the reaction. He$^+$ projectiles are dumped
in a Faraday cup and only He$^{2+}$ particles
from the electron loss reaction are measured with
a second position sensitive detector. They serve
as a trigger for our coincidence experiment. The
data are acquired and stored in list mode format
event-by-event. The typical times-of-flight were
12 $\mu$s for H${_2}^+$ and 17 $\mu$s for
D${_2}^+$. The momentum resolution was determined
by the simultaneously measured electron capture
reaction $He^+ + H_2 / D_2 \rightarrow He^0 +
{H_2}^+ / {D_2}^+$ to be 0.1 a.u. Small fractions
of false coincidences ($<$ 10 $\%$) were
subtracted. The main source for this background
is the single ionization with a wrongly detected
projectile. In momentum space they are located
around $p_{\parallel}$ $\approx$ 0 and in
perpendicular direction constantly from 0 to
maximum (time-of-flight background). Therefore we
took an average of events, being left and right
of the corresponding H$_2$ or D$_2$
time-of-flight peak and subtracted these from the
data (although the difference between left and right side
of the time-of-flight peaks was only $\pm$ 0.3 $\%$).

\vskip 10mm {\bf III. Results and discussion} \vskip 3mm

One possible source of errors in this experiment
is the dissociation of D${_2}^+$ which leads to a
D$^+$ ion that has the same time-of-flight as
H${_2}^+$ and also different detection efficiency
than D${_2}^+$ ($\approx$ 3 $\%$ \cite{Dub05}).
These problems can be solved by normalization of
the $ee$ process of the measured H${_2}^+$ ions
to the D${_2}^+$ ions. Furthermore, to avoid the
possible problem of regions of reduced detection
efficiency on the detector, we changed the
position of the detector and measured the
H${_2}^+$ and D${_2}^+$ ion spots at different
positions on our detector. To avoid background from dissociating D$^+$
and also H${_2}^+$ from the residual gas and to clean the spectra we
analyze only events with positive momentum in jet direction (y) without
loss of generality. In this way we can completely separate the signal of H${_2}^+$ and D$^+$.

For the main results of this paper we choose He$^+$ projectiles with an energy of
650 keV (162.5 keV/u) colliding with H$_2$ and
D$_2$. This projectile energy had to be tuned in
a way that both processes $ee$ and $ne$ can be
clearly resolved. The $ne$ process dominates at
smaller projectile energies, while the $ee$
process is dominant at higher energies. The ratio
of both processes for H${_2}^+$ depending on the
projectile energy can be seen in Fig. 2 and 3.

\begin {figure}[htbp]
  \begin{center}
    \includegraphics[width=1.0\linewidth]{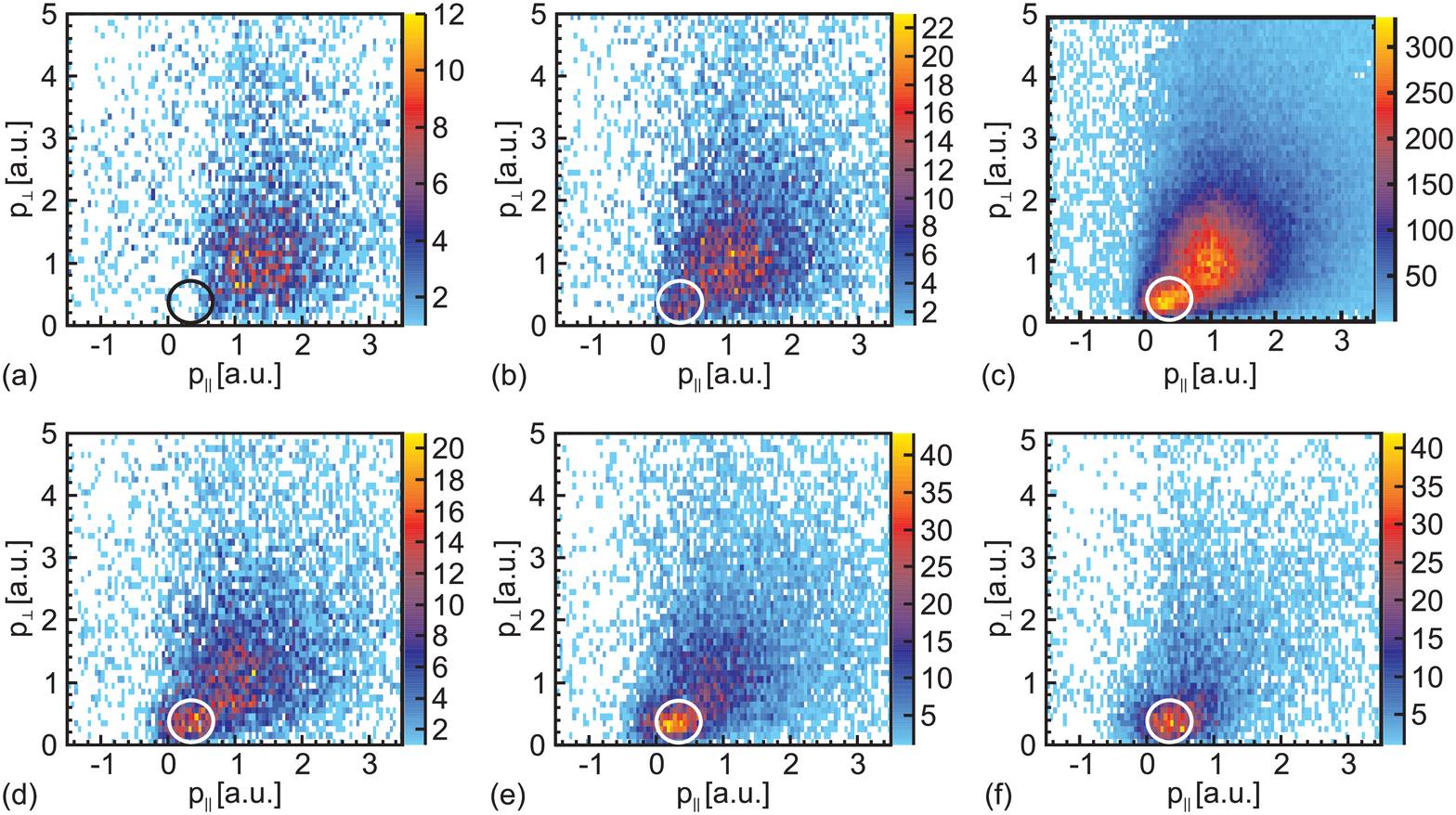}
    \caption{(Color online) H${_2}^+$ recoil ion momenta transverse and longitudinal to the incident
    beam direction for 400 keV (a), 550 keV (b), 650
keV (c), 700 keV (d), 900 keV (e) and 1500 keV
(f) He$^+$ projectile energy (reaction (1)). The $ee$ process is marked by the circle line.}
  \label{fig2}
  \end{center}
\end {figure}

The projectile energy dependence of the $ee$ process is shown in Fig. 3. In the H${_2}^+$ doubly differential cross sections a circle of 0.35 a.u. radius around $p_{\parallel}$ = 0.35 a.u. and $p_{\perp}$ = 0.35 a.u. (circle line in Fig. 2) was selected which characterizes the $ee$ part of all counts in these spectra. The ratio of this part to all counts was plotted and shows an expected rise of the $ee$ process depending on the projectile energy.

\begin {figure}[htbp]
  \begin{center}
    \includegraphics[width=0.66\linewidth]{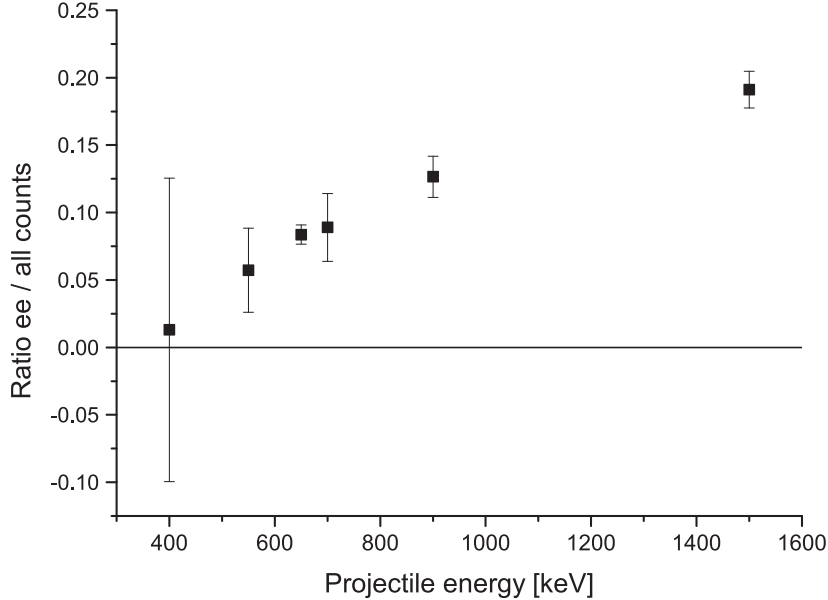}
    \caption{Ratio of the $ee$ process to all counts in mutual projectile and target ionization in He$^+$ + H$_2$ collisions in dependence of the projectile energy.}
  \label{fig3}
  \end{center}
\end {figure}

After determining the $ee$ to $ne$ ratio for different
projectile energies, we chose 650 keV for our
comparison of H$_2$ versus D$_2$, because both
processes are similarly intense at this energy.
Both molecules are separated in time-of-flight as
well as in position on our detector. The recoil ion momenta
perpendicular and parallel to the projectile beam
direction are plotted in Fig. 4. H$_2$ is shown
on the left, D$_2$ on the right. In the region of
small momentum transfers ($p_{\parallel}$ $<$ 0.5
a.u. and $p_{\perp}$ $<$ 1 a.u.) both spectra
show a maximum which corresponds to the $ee$
process. The second maximum at larger total
momenta corresponds to the $ne$ process.

\begin {figure}[htbp]
  \begin{center}
    \includegraphics[width=1.0\linewidth]{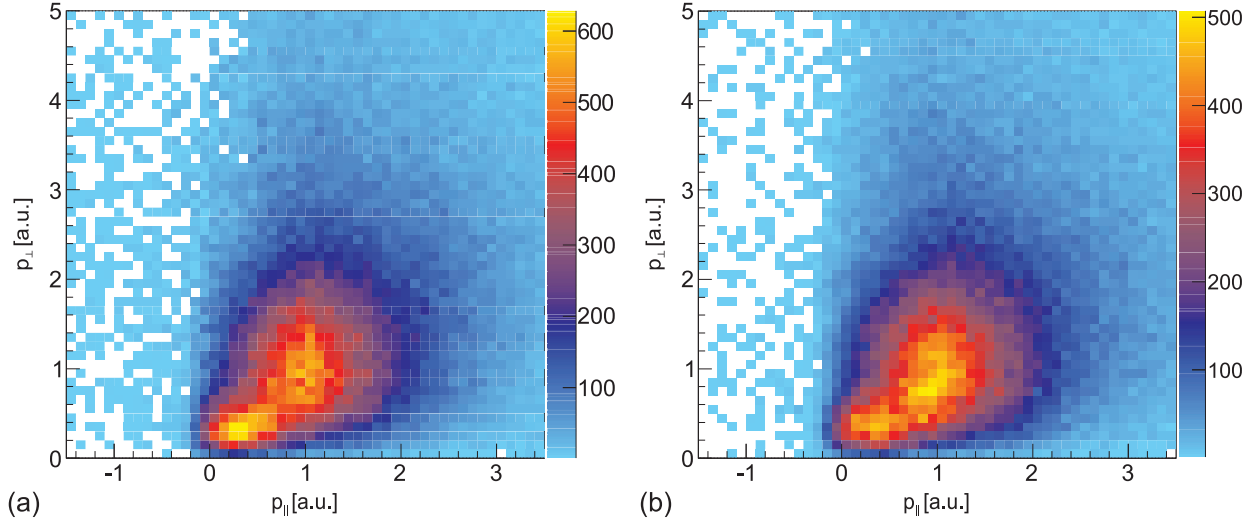}
    \caption{(Color online) Recoil ion momenta transverse and longitudinal to the incident beam direction for mutual projectile and target ionization for 650 keV He$^+$ impact onto H$_2$ (a) and D$_2$ (b).}
  \label{fig4}
  \end{center}
\end {figure}

For a more detailed comparison the longitudinal
and transverse momenta of both isotopes are
presented separately in Fig. 5. The data have
been normalized to the integral (not peak
maximum). In both directions the momentum
distributions are almost identical in shape. The
size of the statistical error bars is smaller
than the points.

\begin {figure}[htbp]
  \begin{center}
    \includegraphics[width=1.0\linewidth]{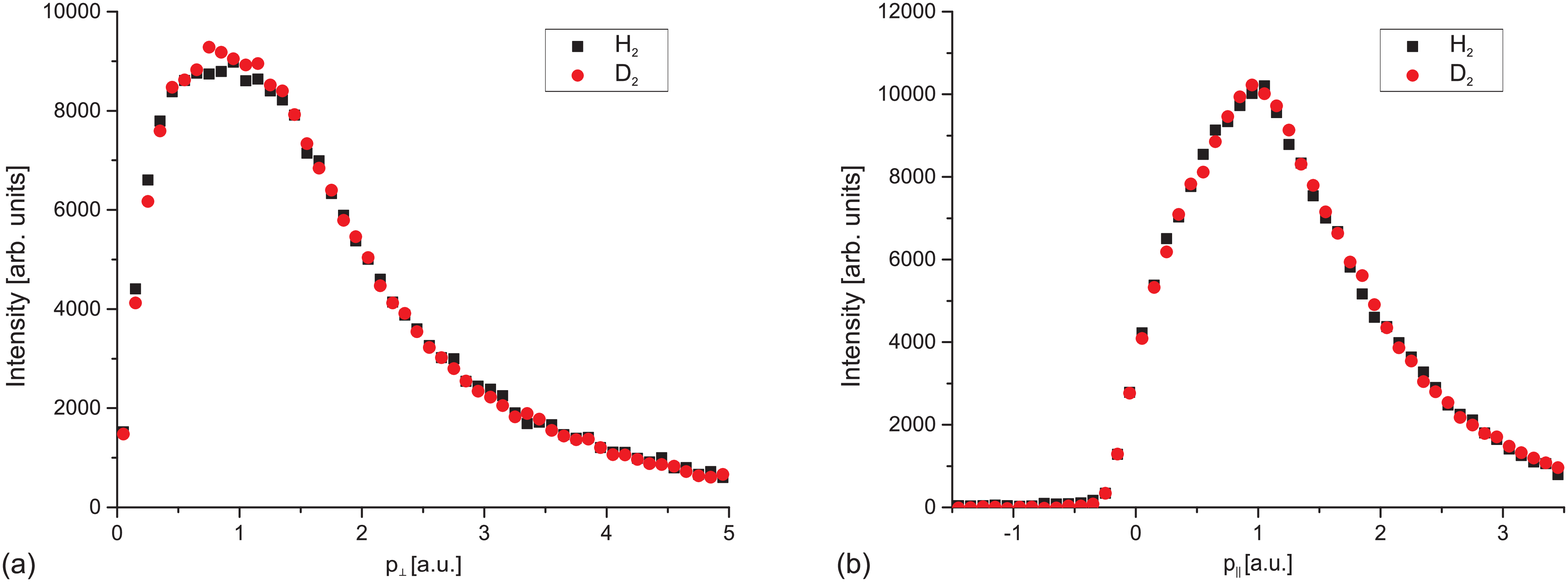}
    \caption{(Color online) Longitudinal and transverse momenta of H${_2}^+$ and D${_2}^+$ for mutual projectile and target ionization by 650 keV He$^+$ impact normalized to their integral. The H$_2$ data points are shown as black squares and the D$_2$ data points as red circles. The data are projections of the data shown in Fig. 4 onto the horizontal or vertical axis.}
  \label{fig5}
  \end{center}
\end {figure}

Fig. 6 shows the ratio of H$_2$
/ D$_2$  for the transverse (a) and the
longitudinal (b) momentum. Small deviations up to
5 $\%$ fluctuate around 0.
Linear fits yield an intercept value of $-0.0077 \pm 0.0047$ and a slope of $-0.0049
\pm 0.0024$ for the case of transverse and an
intercept of $0.0065 \pm 0.0058$ and a slope of
$-0.0065 \pm 0.0039$ for the case of longitudinal
momentum. These small numbers give a good idea of
the diminutiveness of any deviation from the
expectations from standard scattering theory. Most importantly however the observed deviations
from unity are far smaller than 30 \% deviation reported for electron or neutron impact
\cite{Coo08, Cha97, Cha03, Cha05}. As outlined in the introduction an isotope
effect on electron scattering of 30 \% would alter
the $ne$ but not the $ee$ contribution. No such
effect is observed in our data in full agreement
with the expectation of standard scattering
theories.

\begin {figure}[htbp]
  \begin{center}
    \includegraphics[width=1.0\linewidth]{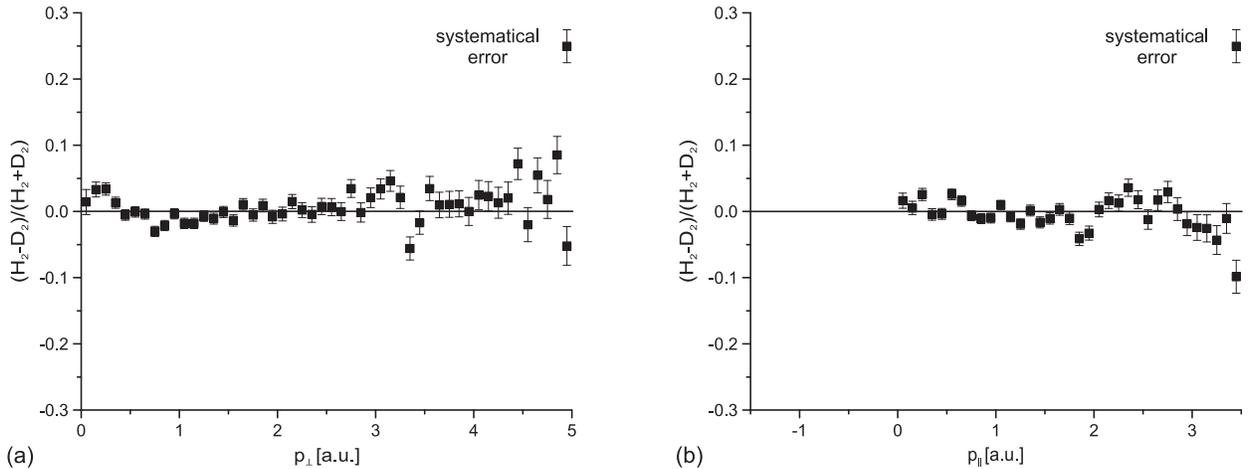}
    \caption{Normalized differences of the H${_2}^+$ and D${_2}^+$ ions (i.e. $(\sigma_{H_2} - \sigma_{D_2}) / (\sigma_{H_2} + \sigma_{D_2})$, results from Fig. 5). In the case of parallel momentum events below 0 a.u. are only background and have no physical meaning, therefore they are not shown.}
  \label{fig6}
  \end{center}
\end {figure}

\vskip 10mm {\bf IV. Conclusions} \vskip 3mm

The goal of the present measurement was to search
for isotope differences in the ionization
dynamics in a heavy particle collision with H$_2$
and D$_2$. This search is motivated by recent
reports on unexpected and so far unexplained
isotope differences in the elastic scattering of
electrons and neutrons \cite{Coo08, Cha97, Cha03,
Cha05} of up to 30 \%. We have investigated
electron loss which for one channel ($ne$) be
thought of as a scattering of a quasi-free electron at
the target nucleus (H$_2$, D$_2$) while a second channel
($ee$) provides an independent in situ
normalization. In contrast to the experimental findings of Cooper et al. the present experimental results do not exhibit any significant differences above 5 \% between the H$_2$ and D$_2$ targets. This 'null' result is in line
with the expectation from all standard scattering
theories. A possible reason for the opposite
conclusion drawn from our experiment as compared
to the surprising results of the electron and
neutron scattering is, that the momentum
transfers are in different regimes: Cooper et al.
\cite{Coo08} used a momentum transfer q of
19.7 a.u., while in our experiment only momentum
transfers of up to 5 a.u. were measured. So a rather different momentum transfer region is probed in this case. At the end of our momentum distribution in Fig. 6 we can see a tendency of an increasing normalized difference for perpendicular momenta and a decreasing normalized difference for parallel momenta. This could be a hint for the reported isotope effect at higher momentum transfers, but experiments that enable these higher momentum transfers are needed to explore this possible momentum transfer dependency, as the observed effect could either depend on the momentum transfer or occur only at higher momentum transfers. Chatzidimitriou-Dreismann suggests an increase of the anomaly with increasing momentum transfer \cite{Cha97, Cha03, Cha05, Cha052}.
For further investigation of this question we plan to perform measurements with higher momentum transfers, but these can not be reached with our Van-de-Graaff accelerator, a storage ring is needed instead.
For the sum of electron energies in the projectile frame, we can here use the approximation $\sum_{i=1,2} E_{kin,e_i}^p \approx p_{\parallel,rec} \cdot v_p = 3.5 a.u. \cdot 2.55 a.u. = 8.93 a.u. \approx 243 eV$ \cite{Doe00}. Although the electron energies are not presented in this paper, we consider only electrons below 243 eV due to 3.5 a.u. maximum longitudinal recoil ion momentum.

\acknowledgments  This work was supported by the DFG and Roentdek Handels GmbH. We acknowledge enlightening discussions with C.A. Chatzidimitriou-Dreismann.\\

\bibliographystyle{srt}
{}

\end{document}